\ifdefined\pdfminorversion\pdfminorversion=7\fi
\documentclass[aps,prl,reprint,superscriptaddress,floatfix,longbibliography,10pt]{revtex4-2}
\usepackage{amsmath,amssymb,amsfonts,bm,graphicx,booktabs}
\usepackage{microtype}
\usepackage{tikz}
\usetikzlibrary{calc}
\usepackage[colorlinks=true,linkcolor=blue,citecolor=blue,urlcolor=blue]{hyperref}
\newcommand{\dd}{\mathrm d}
\newcommand{\ii}{\mathrm i}
\newcommand{\CS}{\operatorname{CS}}
\newcommand{\tr}{\operatorname{tr}}

\newcommand{\IQH}{\mathcal L_{\mathrm{IQH}}[A]}
\newcommand{\bk}{\mathbf k}

\newcommand{\one}{\mathbf1}
\newcommand{\prlsection}[1]{\paragraph*{#1.}}
\begin{document}

\title{Direct transition from a fractional Chern insulator to a topological superconductor}
\author{Taige Wang}
\affiliation{Department of Physics, Harvard University, Cambridge, Massachusetts 02138, USA}
\affiliation{Materials Research Laboratory, Massachusetts Institute of Technology, Cambridge, MA 02139, USA}
\date{\today}

\begin{abstract}
We find numerical evidence for a direct continuous transition from a Laughlin fractional Chern insulator (FCI) to a $c_-=3/2$ topological superconductor (TSC), driven by increasing bandwidth at filling $\nu=1/3$ in a repulsively interacting Chern band. At the critical point, the electron correlation length diverges at momentum away from the reference Fermi surface. We develop a generalized theory of superconducting-vortex proliferation for this transition. Magnetic translation organizes the vortices into six valleys, so translation-preserving proliferation requires additional gauge fields. The resulting theory reproduces the momentum structure of the critical spectrum and captures the divergent electron correlation. Upon doping away from commensurate filling, the theory gives rise to a superconducting dome. These results are consistent with an anyonic origin of the topological superconductivity in the absence of a weak-coupling Fermi-surface instability.
\end{abstract}
\maketitle

Can superconductivity emerge from an anyon fluid rather than from a Fermi-surface pairing instability~\cite{Kohn1965Pairing,Shavit2025Geometry,Jahin2026Geometry,Zou2026KohnLuttinger,XuCheng2025ChiralSC,ChenYinqi2026FiniteMomentum,MayMann2026PairingTopology}? The discovery of fractional Chern insulators (FCIs) in twisted bilayer MoTe$_2$ established fractional quantum Hall physics at zero magnetic field~\cite{Cai2023,Park2023,Zeng2023,Xu2023}. The subsequent observation of superconductivity (SC) near fractional and reentrant integer quantum anomalous Hall phases~\cite{Xu2025FCISC} brings the possibility of anyon superconductivity into experimental focus. Laughlin first proposed that fractional statistics could give rise to superconductivity~\cite{Laughlin1988,LaughlinScience,Fetter1989,Chen1989,WenZee1990,LeeFisher1989,FisherLee1991}. Recent theories find routes to superconductivity by doping fractionalized topological states or pairing fractionalized carriers~\cite{ShiSenthilDoping,Nosov2026,ZhangHolon,Fan2026,ShiSenthilFluid,ZhangColor,Mehta2026,WangU3,WangZaletel,Song2021Doping,TangWen,Kim2025}. Numerical studies have likewise found superconductivity near FCIs, chiral spin liquids, and related topological transitions~\cite{WangZaletel,Pichler2026,Divic2025,Guerci2025,Kuhlenkamp2025}.

In our previous work~\cite{WangZaletel}, we found a chiral $f-if$ superconductor near the $\nu=1/3$ Laughlin FCI in a model of repulsively interacting spin-polarized electrons. Its pairing symmetry identifies the weak-coupling side as $c_-=3/2$, and adiabatic continuity extends this assignment across the superconducting phase. Generalized string order independently establishes its topological nature, supporting the half-integer \(c_-\) assignment and the presence of non-Abelian vortices~\cite{ZhangWangZaletelTopology}. Here we find numerical evidence that increasing bandwidth drives a direct continuous transition from the $c_-=1$ Laughlin FCI to this topological superconductor. At the critical point, the electron correlation length diverges at a momentum away from the reference Fermi surface, while neutral correlations peak on the \(\mathbf G/6\) momentum grid. Toward the metal, the dominant electron correlations in the superconducting phase return to the reference Fermi surface. 

Despite the broad literature on anyon superconductivity, known anyonic routes to the $c_-=3/2$ topological superconductor remain limited to weak pairing of charge-$e/3$ anyons~\cite{WangU3,ZhangColor,Mehta2026,Fan2026,ShiSenthilFluid,HanOrthogonal}. Existing direct-transition theories do not connect the $c_-=1$ Laughlin FCI to this state: our previous work gives a direct transition to a $c_-=3$ chiral superconductor~\cite{WangZaletel,ShiSenthilDisorder,Nosov2026}, while QCD--CS theories lead to an $f+if$ superconductor of opposite chirality~\cite{MaHe,ShiSenthilMelting}. The numerical results further constrain the theory through lattice translation and the momentum structure of the critical correlations.

\begin{figure}[t]
\centering
\includegraphics[width=\columnwidth]{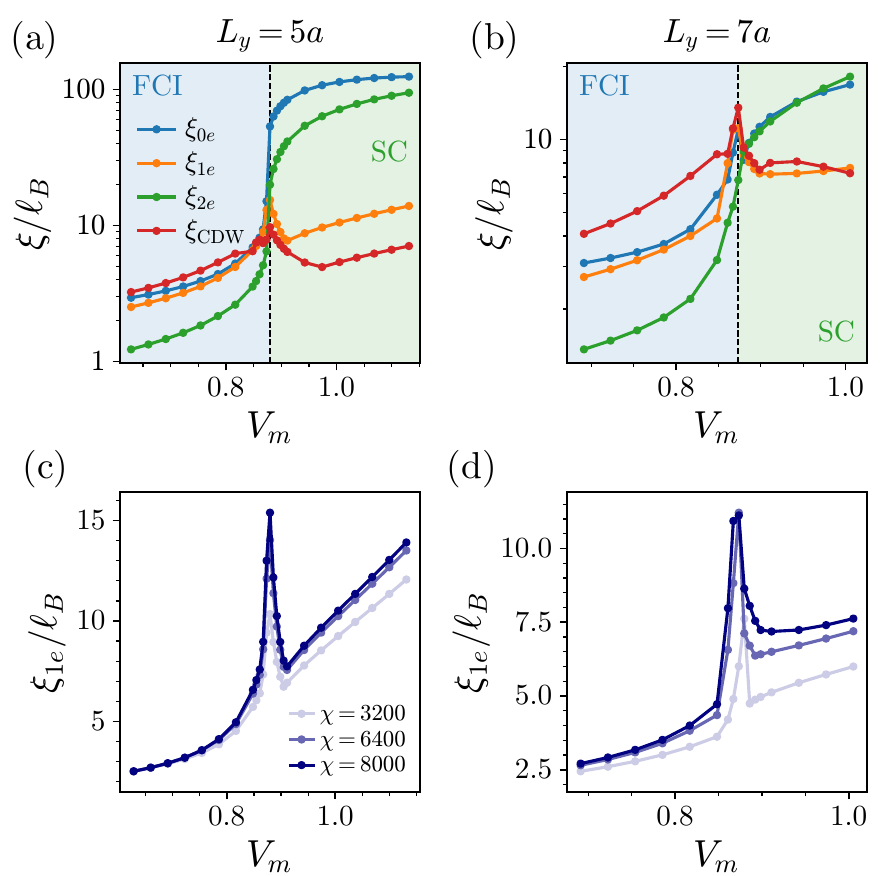}
\caption{Correlation lengths across the FCI--SC boundary. (a,b) Leading neutral, electron, pair, and finite-momentum neutral correlation lengths for $L_y=5a$ and $7a$ at $\chi=8000$. (c,d) Electron correlation length near the transition for multiple bond dimensions. The critical points lie near $V_m=0.88$ and $0.87$ for $L_y=5a$ and $7a$, respectively. The curves labeled CDW show finite-momentum neutral correlation lengths.}
\label{fig:correlations}
\end{figure}

We construct such a theory by proliferating the non-Abelian vortices of the topological superconductor. At $\nu=1/3$, magnetic translation organizes these vortices into six valleys. Translation-preserving proliferation then requires additional statistical gauge fields, and the resulting Higgs phase is the Laughlin FCI. The theory reproduces the momentum structure of the critical electron and neutral correlations. Away from commensurate filling, doping suppresses vortex proliferation and produces a superconducting dome. Together, these results are consistent with superconductivity emerging from an anyon fluid rather than from a Fermi-surface pairing instability.

\begin{figure}[t]
\centering
\includegraphics[width=\columnwidth]{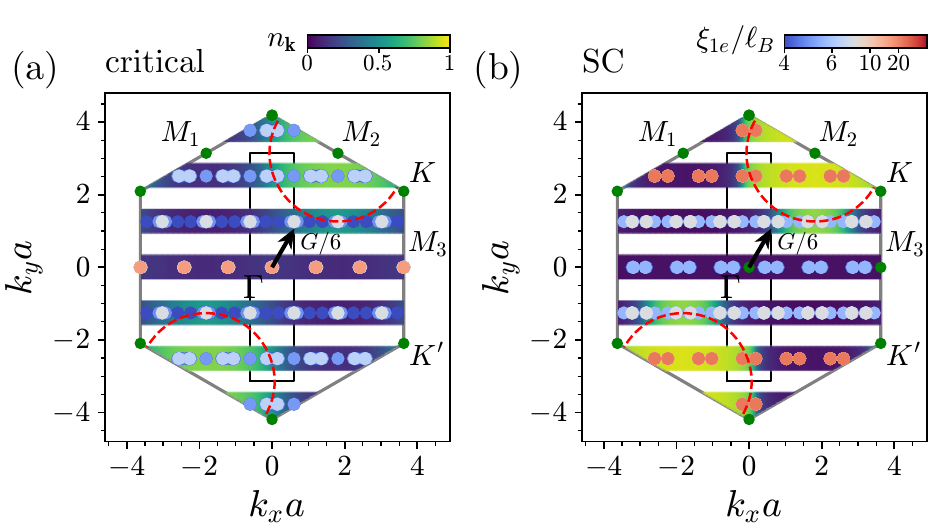}
\caption{Electron momentum structure for $L_y=5a$ at (a) the FCI--SC boundary, $V_m=0.88$, $\chi=8000$, and (b) inside the SC toward the metal, $V_m=1.38$, $\chi=10000$. Colors give $\xi_{1e}/\ell_B$, wires show the momentum-space occupation $n_{\mathbf k}$ of the corresponding state, and dashed curves mark the single-particle reference Fermi surface. Black boxes show the reduced Brillouin zone of the MPS unit cell, $|k_x|\le G_{1x}/12$ and $|k_y|\le\pi/a$. Each measured mode is displayed in six copies related by $k_x\rightarrow k_x+nG_{1x}/6$, with $n=0,\ldots,5$.}
\label{fig:electron-comparison}
\end{figure}

\prlsection{Direct transition between FCI and SC}

We study the particle--hole-conjugate of the model in Ref.~\cite{WangZaletel}, with spin-polarized electrons projected to the lowest Landau level in a triangular periodic potential,
\begin{equation}
\hat H=
\frac12\sum_{\mathbf q}V(\mathbf q):\rho_{-\mathbf q}\rho_{\mathbf q}:
+V_m\sum_{s=1}^{6}\rho_{\mathbf G_s},
\label{eq:Hamiltonian}
\end{equation}
where $\rho_{\mathbf q}$ is the unprojected density operator, $\mathbf G_s$ are the six shortest reciprocal-lattice vectors, and $a$ is the lattice constant of the triangular potential. We use the repulsive $V_1$ Haldane pseudopotential~\cite{Haldane1983} with $2\pi$ magnetic flux through each unit cell. Thus $V_m$ tunes the bandwidth without changing the single-particle Bloch wavefunctions.

We solve the model using infinite-cylinder DMRG with matrix product states (MPS) that conserve charge and transverse momentum~\cite{ZaletelMong2012,ZaletelMongPollmann}. We use bond dimensions up to $\chi=10000$ and focus on $\nu=1/3$ with $L_y=5a$ and $7a$, where both cylinders exhibit a direct bandwidth-tuned FCI--SC transition discussed below. The $L_y=6a$ cylinder instead shows strong competition between SC and a commensurate $\sqrt3\times\sqrt3$ charge-density wave~\cite{WangZaletel}.

We characterize the transition using correlation lengths extracted from the infinite-MPS transfer matrix~\cite{Zauner2015Transfer,Lee2018QED3,WangSong}. The transfer matrix resolves charge sectors, and we denote the leading neutral, electron, and pair correlation lengths by $\xi_{0e}$, $\xi_{1e}$, and $\xi_{2e}$. A finite correlation length indicates a gapped channel, while a divergence signals a critical mode.

Fig.~~\ref{fig:correlations}(a,b) shows a single phase boundary separating the FCI and SC. The pair and neutral correlation lengths increase as expected upon entering the superconductor, while the electron remains gapped in both phases and becomes critical only at the phase boundary. We fit $\xi_{1e}\sim |V_m-V_{m,c}|^{-\nu}$ with $\nu\approx0.4$ and confirm the divergence of $\xi_{1e}$ by finite-entanglement scaling. The single phase boundary and the power-law divergence of $\xi_{1e}$ provide evidence for a direct continuous FCI--SC transition.

We next resolve the correlation lengths in momentum. Following Refs.~\cite{Lee2018QED3,WangSong}, we write $O(\mathbf r)=\sum_{\mathbf k}e^{i\mathbf k\cdot\mathbf r}O_{\mathbf k}(\mathbf r)$, where $O_{\mathbf k}(\mathbf r)$ varies slowly on the scale of the lattice, and denote its charge-$Q$ correlation length by $\xi_Q(\mathbf k)$. As a numerical consequence of the six-row MPS unit cell, the longitudinal real-space period is enlarged by six, so the transfer matrix resolves $k_x$ only modulo $G_{1x}/6$. When unfolded over the full Brillouin zone, each measured mode is repeated in six copies at $k_x+nG_{1x}/6$, $n=0,\ldots,5$.

The momentum of the critical electron further distinguishes the transition from a Fermi-surface pairing instability. The dashed curve in Fig.~~\ref{fig:electron-comparison} is the single-particle reference Fermi surface obtained from the noninteracting band structure. Deeper in the superconductor toward the metal, the momentum-space occupation $n_{\mathbf k}$ remains smooth but varies most strongly near this reference Fermi surface, and the longest electron correlation is found only at the corresponding Fermi-surface crossings. At the critical point, by contrast, the critical electron lies on the $k_y=0$ row away from the reference Fermi surface. The critical electron is therefore not a Fermi-surface mode.

\begin{figure*}[t]
\centering
\includegraphics[width=\textwidth]{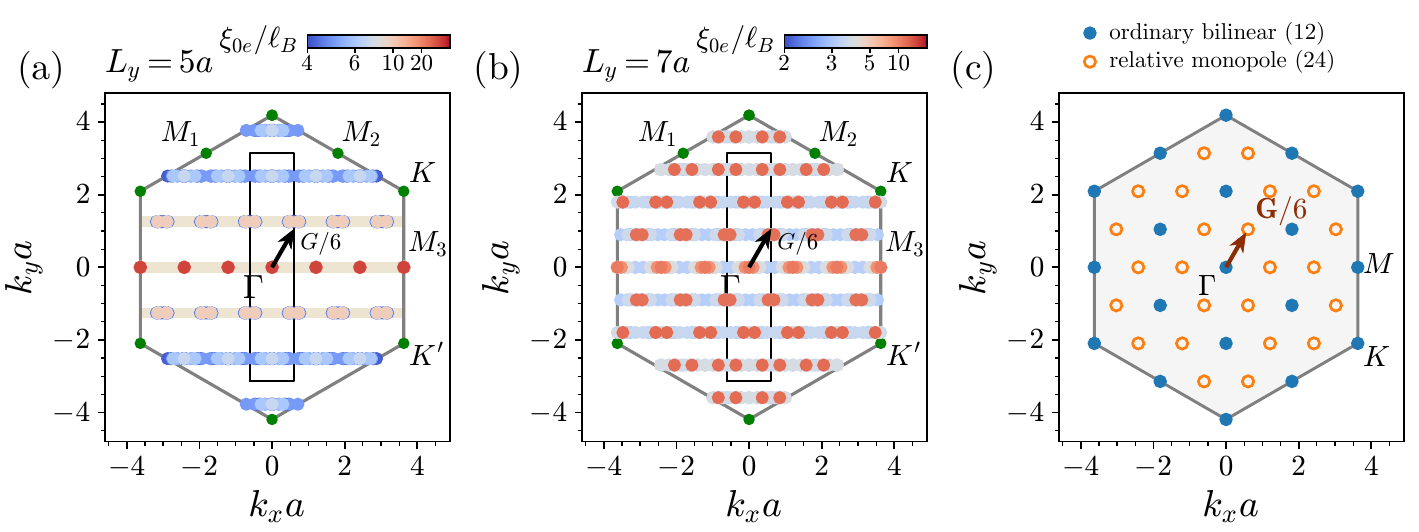}
\caption{Neutral critical momentum structure. (a,b) Neutral correlation lengths at the transition for $L_y=5a$ and $7a$, at $V_m=0.88$ and $0.87$, respectively, with $\chi=8000$. Colors show $\xi_{0e}/\ell_B$. Black boxes shows the reduced Brillouin zone as in Fig.~\ref{fig:electron-comparison}. (c) The vortex theory predicts the operator momenta $\mathbf{k}_{pq}=(p\mathbf G_1+q\mathbf G_2)/6$, with $p,q\in\mathbb Z_6$. Filled symbols denote ordinary vortex bilinears, while open symbols require relative-monopole operators.}
\label{fig:BZ}
\end{figure*}

In addition to the expected zero-momentum Goldstone mode, the neutral correlations show prominent finite-momentum peaks at the critical point. These peaks are compatible with the $\mathbf G/6$ grid shown in Fig.~~\ref{fig:BZ}, although the finite cylinders do not resolve the full two-dimensional grid.\footnote{For $L_y=ma$, $k_ya=2\pi n/m$. The $m=5$ and $7$ cylinders therefore sample different $k_y$, neither exactly at $k_y=G_{1y}/6$ but both nearby.} Together with the off-Fermi-surface critical electron, these neutral peaks reveal a nontrivial momentum structure of the critical field theory.

\prlsection{Non-Abelian vortex proliferation}

We now develop a non-Abelian vortex-proliferation theory for the observed SC--FCI transition. For a bosonic superfluid at integer filling, condensing the elementary vortex restores charge conservation and produces a trivial insulator~\cite{LeeFisher1989,FisherLee1991,Peskin1978,DasguptaHalperin1981,FisherLee1989Duality,SenthilFisher2000}. At fractional filling, however, a gapped charge-conserving descendant cannot be both featureless and translation symmetric, as required by the filling constraint~\cite{Balents2005,ChengSET,LuRanOshikawa}.

This obstruction is encoded directly in the magnetic translation algebra of the vortices. At electron filling $\nu$, the Cooper-pair density is $\nu/2$ per unit cell. An $hc/2e$ vortex therefore sees dual flux $\Phi_{\rm dual}=\pi\nu$ per unit cell, giving
\begin{equation}
 T_1T_2=e^{\ii\pi\nu}T_2T_1 .
 \label{eq:magnetic}
\end{equation}
At $\nu=1/3$, the minimal magnetic multiplet therefore contains six vortex valleys.

In addition, the $hc/2e$ vortices of the $c_-=3/2$ TSC are non-Abelian and carry an odd Majorana core parity~\cite{ReadGreen,Kitaev}. Therefore, we represent the six vortex valleys by $U(2)$-fundamental bosons $\varphi_v$, with $v\in\mathbb Z_6$. Write the $U(2)$ connection as
$\alpha=a+\alpha_0\one_2$, where $a$ is traceless and $\alpha_0=\tr\alpha/2$. The vortices carry unit charge under $\alpha_0$, giving the $hc/2e$ vorticity, and transform as $SU(2)_2$ fundamentals, accounting for the Majorana mode at the vortex core. Define $\IQH=A\dd A/(4\pi)+\Omega_g$ and
$\CS[a]=\tr(a\dd a-2\ii a^3/3)/(4\pi)$. The vortex theory is~\cite{FradkinPf,FradkinLG,ClarkeNayak,GoldmanLG}
\begin{equation}
\begin{aligned}
 \widetilde{\mathcal L}={}&
 \sum_{v=0}^{5}|D_{a+\alpha_0\one_2}\varphi_v|^2
 +V_\varphi+\mathcal L_{\rm Maxwell}\\
 &+2\CS[a]
 +\frac2{2\pi}A\dd\alpha_0
 +3\IQH ,
\end{aligned}
\label{eq:bare-vortex}
\end{equation}
where $\mathcal L_{\rm Maxwell}=(\dd\alpha_0)^2/(4g^2)$. The trace photon represents the superconducting Goldstone mode.

The six fundamental vortex fields cannot condense symmetrically. Assemble the six color doublets into the $2\times6$ matrix
$\widetilde F=(\varphi_0,\ldots,\varphi_5)$. Since the magnetic translation representation is irreducible, translation symmetry requires
\begin{equation}
 \widetilde F^\dagger\widetilde F=\rho\,\one_6 .
 \label{eq:rank-condition}
\end{equation}
For $\rho>0$, the right-hand side has rank six, while
$\operatorname{rank}(\widetilde F^\dagger\widetilde F)\le2$. A fundamental-Higgs condensate of Eq.~\eqref{eq:bare-vortex} therefore necessarily breaks translation.

We resolve this obstruction by dividing the six valleys into three pairs. Write $v=(i_v,\tilde v)$, where $i_v\in\mathbb Z_3$ labels the pair containing valley $v$ and $\tilde v\in\mathbb Z_2$ labels the two valleys within each pair, and couple each pair to an independent $U(1)$ gauge field $\gamma_{i_v}$. Let $\Phi_v$ be $U(2)$-fundamental bosons carrying charge $-1$ under $\gamma_{i_v}$. The physical vortex is then the dressed field
\begin{equation}
 f_{v,a}=\eta_{i_v}\Phi_{v,a},
 \label{eq:vortex-factorization}
\end{equation}
where $\eta_i$ is a fermionic field carrying unit charge under $\gamma_i$. The symmetry fractionalization of $f$ reconstructs the six-valley magnetic translation algebra in Eq.~\eqref{eq:magnetic}. The particle--hole-related $\nu_{\rm e}=2/3$ description admits a simple electron-parton realization
$c=\epsilon_{ab}b_af_b$. Each class now contains two valleys and two color directions, allowing all six modes to condense without the rank obstruction above.

Representing the $3\IQH$ sector by the three level-one fields $\gamma_i$ gives the critical theory
\begin{equation}
\begin{aligned}
 \mathcal L={}&
 \sum_v|D_{a+(\alpha_0-\gamma_{i_v})\one_2}\Phi_v|^2
 +V_\Phi+\mathcal L_{\rm Maxwell}+2\CS[a]\\
 &-\frac1{4\pi}\sum_i\gamma_i\dd\gamma_i
 +\frac1{2\pi}A\dd
 \left(2\alpha_0-\sum_i\gamma_i\right).
\end{aligned}
\label{eq:action}
\end{equation}
The transition is tuned by the mass
$r\sum_v|\Phi_v|^2$ in $V_\Phi$.

When $\Phi$ is gapped, integrating out the statistical connections gives
\begin{equation}
 \mathcal L_{\rm SC}
 =2\CS[a]
 +\frac2{2\pi}A\dd\alpha_0
 +3\IQH
 +\mathcal L_{\rm Maxwell},
 \label{eq:SC}
\end{equation}
recovering the original $c_-=3/2$ superconducting theory.

When $\Phi$ condenses, write
$F_i=(\Phi_{i0},\Phi_{i1})$ as a $2\times2$ matrix, with color as the row index and valley as the column index. A full-rank Higgs condensate is
\begin{equation}
 \langle F_i\rangle=\sqrt\rho\,\one_2,
 \qquad i=0,1,2 .
 \label{eq:full-Higgs}
\end{equation}
Because $F_i$ is gauge charged, translation symmetry only requires the translated condensate to be gauge equivalent to itself. Let $X$ and $Y$ denote Pauli matrices acting from the right in valley space, while $\boldsymbol\sigma$ acts from the left in color space. We denote the action of translations on $\Phi$ by $\widetilde T_{1,2}$. In a convenient gauge,
\begin{equation}
\begin{aligned}
 \widetilde T_1:\quad
 F_i&\mapsto
 e^{-\ii\pi/2}(\ii\sigma_x)F_{i+1}X,
 &\gamma_i&\mapsto\gamma_{i+1},\\
 \widetilde T_2:\quad
 F_i&\mapsto
 (\ii\sigma_y)F_{i-1}(-\ii Y),
 &\gamma_i&\mapsto\gamma_{i-1}.
\end{aligned}
\label{eq:Higgs-translations}
\end{equation}
The factors acting from the left are the compensating $U(2)$ gauge transformations. Equation~\eqref{eq:full-Higgs} is invariant,
\begin{equation}
 \widetilde T_1\langle F_i\rangle
 =
 \widetilde T_2\langle F_i\rangle
 =
 \sqrt\rho\,\one_2 .
 \label{eq:Higgs-invariance}
\end{equation}
The $C_3$ symmetry is preserved in the same way.

The condensate Higgses the $SU(2)$ color sector,
\begin{equation}
 \gamma_0=\gamma_1=\gamma_2=\alpha_0 .
 \label{eq:gauge-locking}
\end{equation}
The remaining common $U(1)$ is described by the Laughlin FCI action
\begin{equation}
 \mathcal L_{\rm FCI}
 =-\frac3{4\pi}\alpha_0\dd\alpha_0
 -\frac1{2\pi}A\dd\alpha_0 ,
 \label{eq:FCI}
\end{equation}
with Hall conductivity $e^2/(3h)$ and $c_-=1$.

The Higgs condensate also makes the fate of the superconducting vortex transparent. Since $f_v=\eta_{i_v}\Phi_v$, condensation of $\Phi$ screens the $\Phi$ factor and identifies $f_v\sim\eta_{i_v}$. Because $\eta_i$ carries one unit of $\gamma_i$ gauge charge, $f$ is further identified with the fundamental Laughlin anyon. The superconducting vortex thus becomes a Laughlin anyon across the transition~\cite{ShiSenthilDisorder,Nosov2026}.

\prlsection{Critical operators and momentum structure}

Local electron operators are gauge-invariant combinations of monopoles and the critical fields $\Phi$. In a Cartan gauge $a=a_3\sigma_z$, write
$\mathcal M_{(m_1,m_2;\mathbf n)}$ for a monopole carrying $2\pi(m_1,m_2)$ flux in the two eigenvalues $\alpha_0\pm a_3$ and $2\pi n_i$ flux in $\gamma_i$. Its electric charge is
\begin{equation}
 \frac{Q}{e}=m_1+m_2-\sum_i n_i .
 \label{eq:monopole-charge}
\end{equation}

The local electron operators are
\begin{equation}
 \mathcal C_{vw}\sim
 \mathcal M^{A}_{(1,0;\mathbf e_{i_v}-\mathbf e_{i_w})}
 \Phi_v^\dagger\sigma^A\Phi_w .
 \label{eq:electron-operator}
\end{equation}
Because of the level-2 $SU(2)$ Chern--Simons term, the $(1,0)$ monopole transforms as an $SU(2)$ triplet, with $A=1,2,3$. Contracting this index with the triplet bilinear $\Phi_v^\dagger\sigma^A\Phi_w$ gives a gauge-invariant local electron operator. Its lowest allowed dressing has half-integer angular momentum and fermionic statistics. Thus a local electron can become gapless at the critical point without an underlying Fermi surface. This sharply contrasts with the previously proposed FCI--$c_-=3$ SC transition~\cite{WangZaletel}, where the critical fields carry charge $2e/3$ and the physical electron remains gapped.

Neutral operators are
\begin{equation}
 \mathcal O_{vw}=
 \mathcal M_{(0,0;\mathbf e_{i_v}-\mathbf e_{i_w})}
 \Phi_v^\dagger\Phi_w ,
 \label{eq:neutral-operator}
\end{equation}
with no monopole when $i_v=i_w$. Thus 12 channels are ordinary bilinears and the remaining 24 require relative statistical flux. Their momenta follow from the six-valley magnetic translation algebra. The definite-momentum combinations are
\begin{equation}
 \mathcal O_{pq}
 =
 \sum_{v,w}
 \left(T_1^qT_2^{-p}\right)_{wv}
 \mathcal O_{vw},
 \qquad
 \bk_{pq}
 =
 \frac{p\mathbf G_1+q\mathbf G_2}{6}.
 \label{eq:operator-momentum}
\end{equation}
where $p,q\in\mathbb Z_6$. The 12 ordinary neutral bilinears occupy $p+q=0\pmod3$, while the 24 relative-monopole channels fill the remaining points, giving the complete $\mathbf G/6$ mesh in Fig.~~\ref{fig:BZ}(c).

\prlsection{Relation to other phases and transitions}

Table~~\ref{tab:phases} summarizes several phases obtained from different states of the vortex matter. A natural competing Higgs state is one in which the condensed vortex spinors are proportional to a single color vector,
\begin{equation}
 \langle\Phi_{i0}\rangle=\sqrt\rho\,u_i,
 \qquad
 \langle\Phi_{i1}\rangle=0,
 \label{eq:collinear-Higgs}
\end{equation}
with $u_0=u_1=(1,0)^T$ and $u_2=(0,1)^T$. Unlike the full-rank condensate, this state selects one valley within each group and breaks translation symmetry. The resulting Higgs phase is a topologically trivial CDW with $c_-=0$.

\begin{table}[!b]
\caption{Vortex phases and the corresponding electronic phases. The listed condensates are gauge-fixed representatives. The collinear Higgs state is given in Eq.~\eqref{eq:collinear-Higgs}. The bIQH row requires additional vortex dynamics beyond the fundamental-Higgs transition.}
\label{tab:phases}
\begin{ruledtabular}
\setlength{\tabcolsep}{2pt}
\begin{tabular}{@{}p{0.49\columnwidth}p{0.49\columnwidth}@{}}
Vortex matter & Electronic phase\\ \hline
Mott, $\langle\Phi_v\rangle=0$
& $\mathrm{SC},\ \sigma_{xx}^{\rm dc}=\infty,\ c_-=3/2$\\[4pt]
Full Higgs, $\langle F_i\rangle=\sqrt\rho\,\one_2$
& $\mathrm{FCI},\ \sigma_{xy}=\frac{1}{3}\frac{e^2}{h},\ c_-=1$\\[4pt]
Collinear Higgs, Eq.~\eqref{eq:collinear-Higgs}
& $\mathrm{CDW},\ \sigma_{xy}=0,\ c_-=0$\\[4pt]
bIQH, $\langle\Phi_v\rangle=0$
& $\overline{\mathrm{FCI}},\ \sigma_{xy}=-\frac{1}{3}\frac{e^2}{h},\ c_-=-1$
\end{tabular}
\end{ruledtabular}
\end{table}

The Abelian counterpart of our construction is dual to the QED--CS theory of the $c_-=3$ superconducting descendant~\cite{Fan2026,WangU3}. In the particle--hole-conjugate $\nu=2/3$ description, the latter is a Chern-number-changing transition of the fermionic parton $f$. By the standard $2+1$D bosonization duality~\cite{KarchTong,SeibergSenthilWangWitten}, each fermionic parton can be written as $f=\eta\Phi$ and the fermion mass changing sign maps to condensation of $\Phi$ with $\eta$ in a fixed $C=+1$ IQH band. Thus the QED--CS transition is the Abelian version of our theory. Replacing $\Phi$ by the $U(2)$ doublets gives the non-Abelian construction above.

The bosonic integer quantum Hall (bIQH) state~\cite{SenthilLevin2013,GroverVishwanath} in Table~~\ref{tab:phases} provides another natural candidate of the vortex matter. The electronic Chern band carries one flux quantum per unit cell, so the superconductor contains two physical vortices per unit cell, giving a vortex density of $2/3$ in each of the three groups. In the bIQH state, the Chern--Simons Gauss laws give $\Phi_{\alpha_0-\gamma_i}=2\pi/3$ per unit cell in each group. Each group is therefore at bosonic filling $\nu_\Phi=2$, naturally favoring a bIQH state with $\sigma_{xy}^{\Phi}=2$ for the $\alpha_0-\gamma_i$ response. The corresponding Mott--bIQH transition has a $U(2)$ QCD--CS description~\cite{MaHe,ShiSenthilMelting}. Fermionizing the vortex field as $\Phi_{i,a}=h_i\xi_{i,a}$, the transition is described by mass inversion of three $U(2)$-fundamental Dirac fermions $\xi_i$. In the parton construction $c=bf$, these two branches correspond respectively to the $c_-=3/2$ TSC and the $\overline{\mathrm{FCI}}$ with $c_-=-1$, so their transition is described by the same QCD--CS theory~\cite{MaHe,ShiSenthilMelting}. Our fundamental-Higgs transition instead connects the $c_-=1$ Laughlin FCI directly to the $c_-=+3/2$ TSC.

\begin{figure}[!t]
 \centering
 \includegraphics[width=0.65\columnwidth]{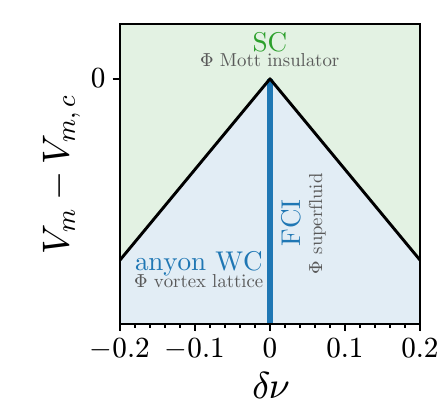}
\caption{Electron phase diagram as a function of $\delta\nu=\nu-1/3$ and $V_m-V_{m,c}$. In the dual vortex description, the vortex-Mott phase is the superconductor, while the vortex superfluid at $\delta\nu=0$ is the FCI. Doping shifts the Higgs boundary and introduces charged flux defects into the vortex superfluid. Their crystallization gives an anyon Wigner crystal (WC).}
 \label{fig:doping}
\end{figure}

\prlsection{Finite anyon density}
Doping away from the FCI introduces a finite density of charged anyons. In the vortex theory, $\delta\nu=\nu-1/3$ appears as additional magnetic flux of $\alpha_0$. The commensurate flux is already encoded in the six vortex valleys, and doping generates an additional field $B$ with
\begin{equation}
 B A_M=\pi\delta\nu .
 \label{eq:addedflux}
\end{equation}
Here $A_M$ is the unit-cell area. A magnetic field moves the bosonic Mott--superfluid transition toward the Mott side by frustrating boson condensation~\cite{Niemeyer1999,Oktel2007,Umucalilar2007,Goldbaum2008}. At mean-field level, the critical mass shifts by
\begin{equation}
 \Delta r_c(B)=-\kappa|B| ,
 \label{eq:finite-density-boundary}
\end{equation}
where $\kappa$ is the vortex stiffness. Translated back to the electronic phase diagram, the Mott--superfluid boundary of $\Phi$ gives rise to the superconducting dome shown in Fig.~~\ref{fig:doping}.

\prlsection{Discussion}
We numerically find a direct transition from the $\nu=1/3$ Laughlin FCI to a chiral TSC with $c_-=3/2$ and construct a non-Abelian vortex-proliferation theory consistent with the observed critical modes. At fractional filling, translation-preserving vortex proliferation requires the additional gauge fields $\gamma_i$. This provides a route for a TSC to emerge directly from a fully gapped FCI without a nearby electron Fermi surface, providing strong evidence that the superconductivity originates from the fractionalized degrees of freedom of the FCI.

Twisted bilayer MoTe$_2$ provides the most direct experimental setting, where superconductivity and fractional quantum anomalous Hall states occur within the same moiré Chern band~\cite{Xu2025FCISC}. Rhombohedral pentalayer graphene provides another setting~\cite{Lu2024FQAH,Han2025ChiralSC}. The critical electron provides a direct experimental test. Scanning tunneling spectroscopy (STS) can map $dI/dV(V,g)$ across the transition~\cite{Liu2026MoTe2STM}, where $g$ is the experimental tuning parameter and $g_c$ its critical value. The tunneling gap edge $V_*$ vanishes as
\begin{equation}
eV_*\sim |g-g_c|^{z\nu} .
\end{equation}
Here we find $z=1$ from the critical theory and $\nu\simeq0.4$ from numerical calculation. Quantum twisting microscopy can additionally resolve the single-electron spectrum in both energy and momentum~\cite{Inbar2023QTM,Waschitz2026QTM}, directly testing our prediction that the gap closes at isolated critical momenta rather than along a Fermi surface. Observing both the gap-closing scaling and isolated critical electron momenta would distinguish the direct FCI--TSC transition from a metallic route to superconductivity.

\begin{acknowledgments}
\textbf{Acknowledgments.} We thank T. Senthil, Michael P. Zaletel, and Ashvin Vishwanath for helpful discussions. We especially thank Zhengyan Darius Shi for discussion on the $\nu = 1/3$ scenario. T.W. is grateful for support from the Harvard Quantum Initiative Fellowship, the Simons Collaboration on Ultra-Quantum Matter, a grant from the Simons Foundation (Grant No. 651440), and the U.S. Department of Energy, Office of Basic Energy Sciences, under Grant No. DE-FG02-03ER46076.
\end{acknowledgments}

\bibliography{arxiv}
\end{document}